# Title

OPRD-Imp: a dataset of experimentally observed impurities and side products in multi-step pharmaceutical syntheses

# Authors

Heejeong Kim[1], Seonbin Kim[2], Jihwan Kim[2], Tim Pinkhassik[3], Magdalena Lederbauer[3], Suong Tran[4], Connor W. Coley[3,5], Juyoung Yoon[1] *, Joonyoung F. Joung[2]*

**Affiliations**

1. Department of Chemistry and Nanoscience, Ewha Womans University, Seoul, 03760, South Korea
2. Department of Chemistry, Kookmin University, Seoul, 02707, South Korea
3. Department of Chemical Engineering, Massachusetts Institute of Technology, Cambridge, MA, USA
4. Department of Chemistry, Massachusetts Institute of Technology, Cambridge, MA, USA
5. Department of Electrical Engineering and Computer Science, Massachusetts Institute of Technology, Cambridge, MA, USA

corresponding author(s): Juyoung Yoon (jyoon@ewha.ac.kr), Joonyoung F. Joung (jjoung@kookmin.ac.kr)

# Abstract

Predicting reaction impurities is a critical yet persistently data-limited challenge in pharmaceutical process chemistry. While machine learning models for reaction outcome prediction have advanced rapidly, benchmarking their ability to predict impurities remains impractical due to the absence of curated, publicly accessible datasets. We present OPRD-Imp, a dataset of 229 organic reactions drawn from 62 publications in Organic Process Research & Development, each annotated with experimentally observed impurities and their mechanistic origins. A total of 784 impurity entries is provided, classified into four origin categories. Reactions are represented as reaction SMILES and impurities as individual SMILES strings, with multi-step synthetic route connectivity preserved. All data were manually extracted from the source publications by trained chemists following a standardized extraction protocol, with expert chemical validation and manual structure drawing. OPRD-Imp is intended to serve as a benchmark resource for evaluating models that predict reaction impurities, providing experimentally reported impurity annotations and mechanistic origin labels that are not systematically represented in conventional reaction datasets.

# Background & Summary

Identifying and controlling impurities is one of the central challenges in synthetic chemistry, from early-stage laboratory research to large-scale pharmaceutical manufacturing. Unanticipated impurities complicate purification, compromise yields, and obscure structure–activity relationships by introducing uncharacterized compounds into biological assays.[1,2] Critically, their formation is difficult to anticipate even for experienced chemists. Side products can arise from competing or cascading pathways, trace metal or catalyst residues, and impurities already present in the starting materials that are carried through and chemically transformed.[3,4] As a result, chemists are frequently confronted with unexpected products and must resort to laborious analytical work to detect and assign them. Structures are typically confirmed either by isolating the impurity on scale-up for NMR characterization, or by forming a structural hypothesis from mass spectrometry and mechanistic reasoning and comparing it against an independently synthesized authentic standard.[5] This makes exhaustive impurity mapping resource-intensive even at the bench scale.

These challenges are amplified as synthetic routes progress toward process development and manufacturing, where incomplete impurity knowledge creates critical blind spots during scale-up and intersects with strict regulatory obligations. Comprehensive impurity profiling is mandatory at every stage of drug development, as governed by the International Council for Harmonisation (ICH) guidelines Q3A, Q3B, and M7, which control impurities, including genotoxic impurities, down to the parts-per-million level.[6–8] Failure to identify and control impurities can result in delayed regulatory submissions, costly batch reprocessing, or post-market drug recalls. Consequently, the experimental burden of impurity identification and these regulatory demands motivate a complementary, data-driven strategy that learns the patterns of impurity formation directly from accumulated experimental observations.

The field of machine learning for organic chemistry has become increasingly effective for a range of reaction-related tasks, including forward reaction prediction, retrosynthetic planning, and reaction-condition recommendation.[9–18] More recently, mechanistic reaction prediction approaches such as FlowER[19] and Reactron[20] have extended reaction modeling beyond major-product prediction by representing elementary reaction steps. These developments suggest that mechanistic and data-driven models could also be applied to impurity and side-product prediction. However, most existing reaction-prediction benchmarks remain focused on the major intended product, while impurity prediction has received comparatively limited attention.[21–24]

A critical bottleneck is the absence of publicly available datasets containing experimentally reported impurity annotations for reaction-level benchmarking. Large reaction datasets such as the United States Patent and Trademark Office (USPTO),[25] the Open Reaction Database (ORD),[26] and ORDerly[17] provide broad coverage of reaction conditions and major products but do not systematically annotate experimentally observed impurities. Even commercially licensable datasets based on the patent literature or journal articles contain a small fraction of entries reporting multiple products. Consequently, even models designed to predict side products or reaction mechanisms cannot be rigorously evaluated against a curated set of experimentally observed impurities.

We selected Organic Process Research & Development (OPRD) as the primary literature source for curating an impurity-focused dataset for several reasons. First, OPRD publications focus specifically on the development and scale-up of synthetic routes for active pharmaceutical ingredients and pharmaceutical intermediates, where impurity identification is particularly consequential. Second, OPRD papers routinely characterize and report the structures of observed impurities in detail, including analytical evidence, as part of the process development procedure.[27] This contrasts with journals focusing more on academic synthesis or discovery chemistry, where impurities are less systematically reported, and with large patent-derived datasets, where impurity data are typically absent. Third, the pharmaceutical process chemistry context of OPRD publications makes the resulting dataset directly relevant to impurity prediction in real-world synthetic workflows.

Herein, we present OPRD-Imp, a publicly available benchmark dataset of experimentally observed organic reaction impurities from selected Organic Process Research & Development literature. The dataset comprises 229 reactions sourced from 62 publications published between 2002 and 2026, containing 784 annotated impurity entries with four mechanistic origin classifications. Key distinguishing features include (1) experimentally reported and structurally curated impurity annotations; (2) origin labels that enable category-specific evaluation of impurity prediction models; (3) connectivity between sequential reaction steps, together with per-impurity provenance paths that trace each carried-over impurity to the species it originated from, enabling analysis of impurity propagation across synthetic routes; and (4) reaction-class annotation of every product and impurity. The dataset was constructed through structured manual extraction followed by multi-round expert curation and validation against the primary literature. Together, these features make OPRD-Imp a curated benchmark for evaluating models for impurity prediction, designed to prioritise annotation reliability over scale.

## Methods

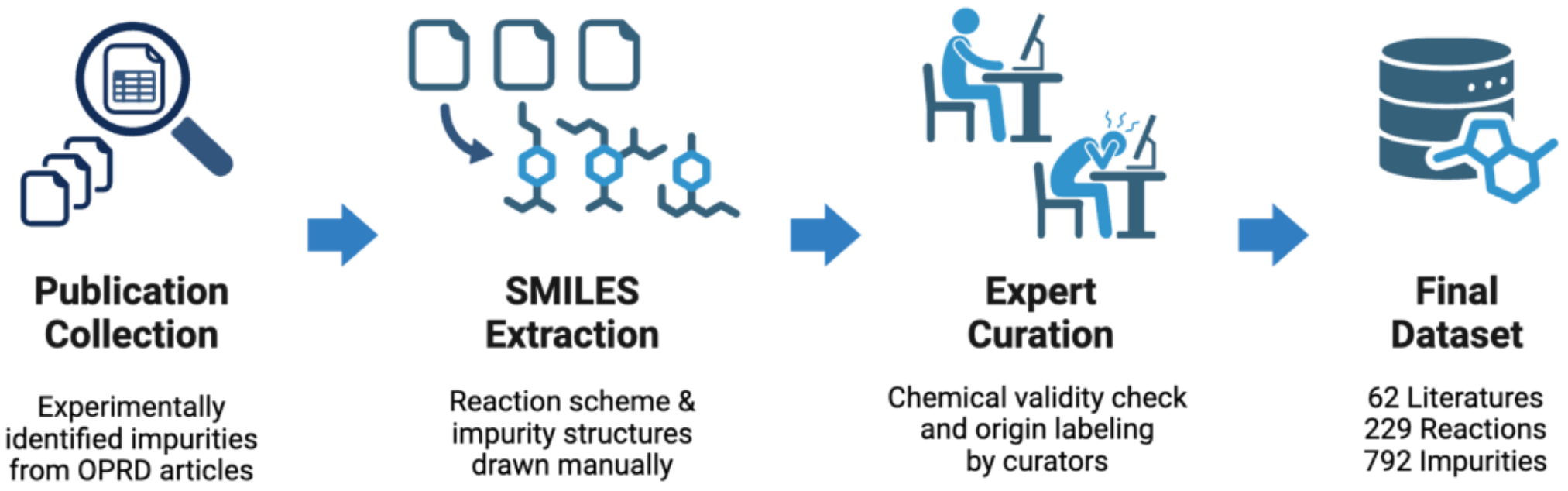

Figure 1. Overview of the dataset construction workflow.

## Data Source and Inclusion Criteria

Publications were identified by searching the Organic Process Research & Development (OPRD) journal website (pubs.acs.org/journal/oprdfk) using the keywords “impurity” or “impurity validation.” All publications published between 2000 and 2026 were considered and the literature search was conducted in May 2026. Articles published before 2000 were not included because, in preliminary screening, they seldom reported impurity structures with supporting analytical evidence and therefore did not meet the inclusion criteria.

Following keyword-based retrieval, each candidate paper was manually reviewed to determine eligibility. We included papers reporting primary experimental data—at least one reaction step performed by the authors, with the reagents, solvents, and reaction conditions specified—in which impurity information is explicitly provided for at least one reaction step, either as a structural assignment or as a description sufficient for unambiguous structure determination. Review articles, perspective pieces, and papers reporting only computational or in silico results were excluded, as were papers in which impurities are mentioned only by name without a clear structural assignment. Following this screening procedure, 62 publications were included in the final dataset.[27–88]

## Structured Information Extraction

The dataset construction workflow is summarized in Figure 1. To ensure systematic and consistent extraction across 62 papers with varying reporting styles, all information was extracted manually from the article PDFs, including the supporting information, by trained chemists following a standardized extraction protocol. Manual extraction without a standardized protocol risks inconsistency between different annotators and inadvertent omission of impurities mentioned only briefly in one location. Therefore, an extraction checklist was designed to provide a systematic framework that can be applied across all papers, reducing annotator-dependent variability.

For each paper, the annotator was required to:
1. Extract the DOI of the paper.

2. Identify and enumerate each reaction step, formatted as "[Starting material].[Reagents] → [Product]" using compound names or numbers given in the paper; these identifiers were converted to SMILES during curation (see Expert Curation and Structure Drawing).
3. Apply an exhaustive impurity extraction checklist, explicitly requiring examination of: abstract, introduction, results and discussion section, experimental section, analytical data, all tables, figures and schemes, and supporting information.
4. For each reaction step, populate a structured table with: compound identifier (name or scheme number as used in the paper), mechanistic origin classification, quantification status, and a mandatory direct quotation from the paper as evidence.

The following categories of impurity report were captured: (i) quantitatively reported impurities (e.g., "X was present at 1.2% by HPLC area"); (ii) qualitatively mentioned impurities (e.g., "trace amounts of compound Y were detected"); (iii) negatively mentioned compounds (e.g., "no formation of Z was observed"); and (iv) implicitly referenced compounds (e.g., "conditions were optimized to suppress formation of W"). This approach was adopted to minimize selective reporting bias in the extraction step itself. Categories (iii) and (iv) were captured so that no compound discussed in the paper was overlooked during screening; however, only compounds that the source reports as actually observed were retained as impurity entries, and compounds explicitly reported as not formed were excluded from the released dataset. Furthermore, a direct supporting quotation from the source text was recorded for every extracted impurity. This provided a concrete anchor for verification in the subsequent curation step. Any impurity lacking a plausible direct quote was flagged for cautious cross-check during expert review.

Each impurity was assigned to one of five origin labels during extraction: the four mechanistic categories defined in Table 1 plus an additional “unknown” label used when the paper does not indicate the origin. Entries labelled “unknown” were resolved into one of the four final categories, or removed, during expert curation, so that the released dataset uses only the four categories in Table 1. The final origin label for every entry was assigned by consensus of the three expert curators; the assignment and verification procedure is described in the Technical Validation section.

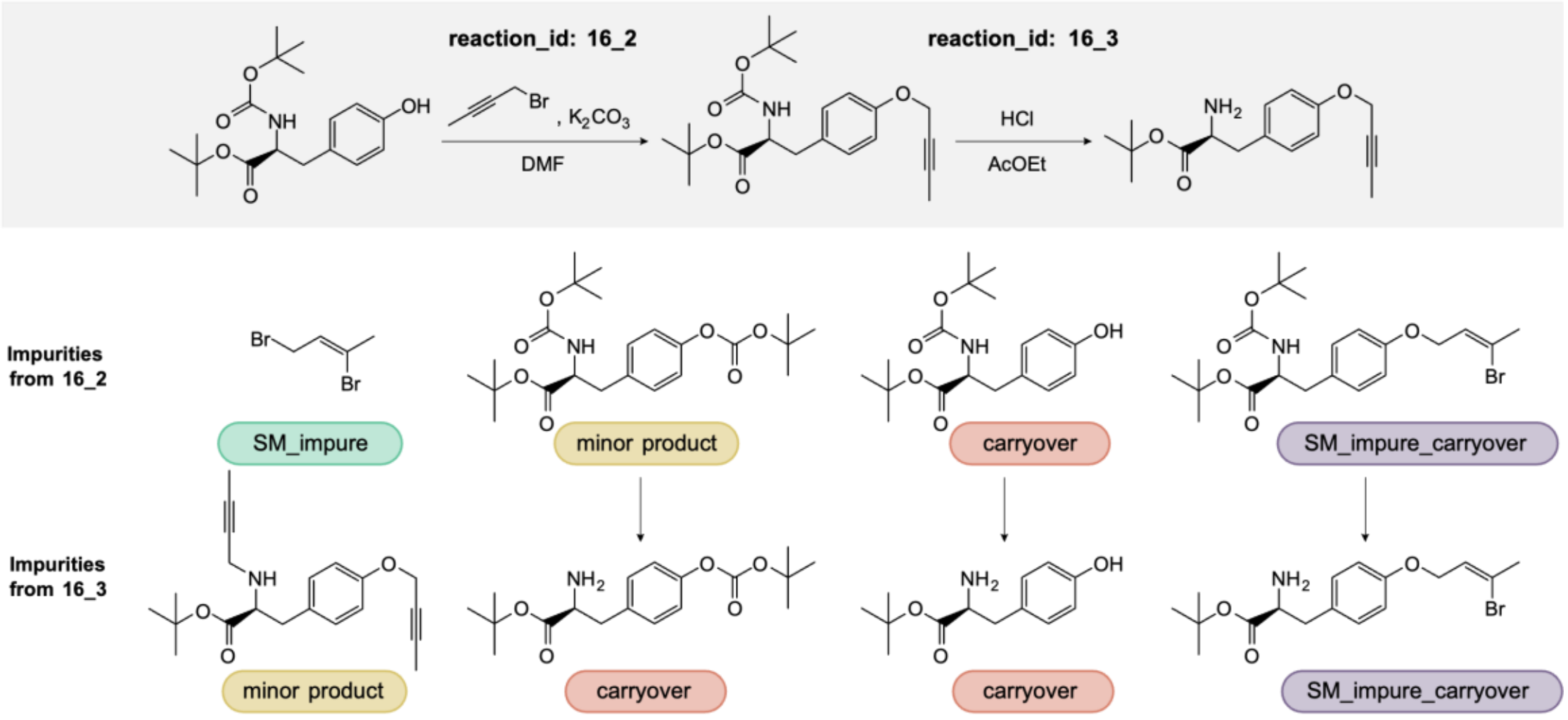

**Figure 2**. Representative multi-step synthetic route illustrating impurity origin classification and cross-step propagation.[27] (a) Two consecutive steps from a six-step synthetic route: 16_2 and 16_3. (b) Impurity origin assignments for each step. Arrows linking impurities across the two steps are drawn for illustrative purposes.

| Origin label | Definition |
|---|---|
| minor_product | An impurity formed through a side reaction during the current reaction step, including regioselectivity or stereoselectivity failures, over-reaction, or other competing transformations. |
| SM_impure | An impurity already present in a starting material or reagent before the current reaction step and not generated by the current reaction. |
| SM_impure_carryover | An impurity originating from an SM_impure species that undergoes chemical transformation during the current reaction step. |
| carryover | An impurity inherited from a preceding synthetic step and transferred into the current reaction without chemical transformation. Unreacted starting material or reagent remaining in the product stream is assigned to this category. |

**Table 1.** Impurity origin categories and their definitions

Origin labels were assigned according to the earliest identifiable source of the impurity relative to the current reaction step: impurities originating from the starting-material feed were classified as SM_impure, their chemically transformed products as SM_impure_carryover, impurities inherited from preceding steps as carryover, and impurities newly formed during the current step as minor_product. Unreacted starting material is recorded only where the source publication explicitly reports it as an impurity in the product stream; such entries are directly identifiable through their parent path referencing a reactant or reagent, so they can be retained or excluded depending on the intended application. Likewise, carryover entries are recorded only for impurities explicitly reported in the source publication.

A representative multi-step synthetic route in which all four origin categories are observed is shown in Figure 2, together with a detailed view of a single reaction step illustrating each category.

The SM_impure category is a special case. These impurities originate from the starting-material feed and are therefore not predictable from the reaction mechanism. Their inclusion nonetheless provides a complete picture of the observable impurity profile as encountered in real process chemistry settings.

## Expert Curation and Structure Drawing

The initial extraction records were reviewed and processed by a team of three trained chemists. Expert curation served four functions.

1. Verification against source documents: Each extracted impurity, together with its recorded evidence quotation, was checked against the original publication, and unsupported or chemically implausible entries were flagged for rejection or correction. The multi-round verification protocol is detailed in the Technical Validation section.

2. Chemical structure drawing and SMILES generation: Each retained impurity, together with the reactants, reagents, solvents, and product of its reaction, was drawn as a chemical structure and converted into a canonical SMILES string, yielding a machine-readable representation of every reaction and its associated impurities. Structures, including stereochemistry, were transcribed as drawn in the source publication. For compounds not explicitly drawn in the source publication (for example, common reagents and solvents referred to only by name), SMILES strings were taken from the corresponding PubChem entries.

3. SMILES representability: Species lacking a standard covalent SMILES representation, such as heterogeneous catalysts and organometallic reagents with coordinate or non-covalent bonds, were encoded using simplified or disconnected ionic conventions, following PubChem representations where available (see Usage Notes).

4. Data structuring and format: Curated data were compiled and structured as JSON to enable structured querying and ease of integration into predictive model evaluation scripts. For papers reporting multi-step synthetic sequences, sequential steps within the same paper were linked using the `next_reaction_id` field. This encodes the directed structure of the synthesis route, allowing downstream users to trace the propagation of carryover impurities through overall steps. Branching routes (where a single intermediate leads to multiple synthetic paths) are represented by providing multiple entries in the next_reaction_id list.

Procedures that the source publication reported as one-pot were retained with the step delineation shown in the paper's reaction scheme, recorded as separate entries linked by `next_reaction_id` and grouped through the `one_pot_group` field, which lists the `reaction_id` keys of all steps in the procedure. Because the intermediates in such procedures were not isolated or independently analysed, observed impurities were attributed to the step at which the source reported them (typically the final step); the empty impurity lists of the preceding one-pot steps therefore do not represent negative examples.

## Data Records

## Repository and File Availability

OPRD-Imp is deposited at figshare[89] and is freely accessible under a Creative Commons Attribution 4.0 International (CC BY 4.0) licence. The repository contains the following files:

- OPRD_imp.json — primary dataset in JSON format (schema described in the JSON Schema section below)
- OPRD_imp_dashboard.html — interactive dashboard for browsing the dataset reaction by reaction, distributed as a single self-contained HTML file (see Usage Notes)

## JSON Schema

The file contains a single top-level JSON object keyed by reaction_id (e.g., "16_2"), so that any reaction can be retrieved directly by its identifier. Each value is a reaction object with the fields described in Table 2.

| Field | Type | Description |
|---|---|---|
| reaction_id (key) | string | Unique identifier in the format N_M (e.g., 16_2), where N (1–62) indexes the source publication and M denotes the reaction step within that publication; used as the key of the top-level object rather than as a field. All eligible reaction steps meeting the inclusion criteria were retained from each selected publication. |
| paper_id | integer | Index of the source publication (1–62); identical to the N component of reaction_id, provided as a separate field for convenient grouping by publication. |
| step | integer | Step number within the synthetic route; identical to the M component of reaction_id. |
| doi | string | DOI of the source publication, recorded on every step of the route. |

| Field | Type | Description |
|---|---|---|
| reaction_smiles | string | Reaction SMILES, reactants.reagents.solvents>>product, assembled from the canonical component SMILES of the fields below. |
| reactants | array of strings | Canonical SMILES of the substrate(s); one entry per reaction, except reaction 26_3, which has two substrates. |
| reagents | array of strings | Canonical SMILES of the reagents; catalysts are recorded as reagents. |
| solvents | array of strings | Canonical SMILES of the solvents; empty array for neat reactions. |
| product | string | Canonical SMILES of the major intended product. |
| name_rxn | string | NameRxn class of the product-forming reaction (see Dataset Statistics). |
| reactants_for_impurities | string | Reactant side used for impurity prediction: the components of reaction_smiles together with every additional species needed to account for the impurities of this reaction, namely the starting-material impurities and the parent species of carried-over impurities. Given as a single dot-separated SMILES string. |
| impurities | array of objects | One object per reported impurity with keys smiles (canonical SMILES), origin (one of the four labels in Table 1), parent (see below), name_rxn (NameRxn class of the impurity-forming reaction), name_rxn_input (the reaction SMILES submitted to NameRxn), and note; empty array for reactions with no reported impurities. |
| notes | string or null | Free-text annotation, mainly documenting non-standard SMILES conventions for heterogeneous catalysts (see Usage Notes); null otherwise. |
| one_pot_group | array of strings or null | reaction_id keys of all steps belonging to the same one-pot procedure, including the entry itself (e.g., ["16_1", "16_2"]); null for reactions that are not part of a one-pot procedure. |
| next_reaction_id | array of strings | Subsequent step(s), each a key of the top-level object; empty array for terminal steps. |

**Table 2.** JSON schema for each reaction entry

For every carryover, SM_impure_carryover, and SM_impure entry, the parent field records the specific species from which the impurity originated: an unconsumed reactant or reagent, or an impurity entry of an earlier step. The value is a path expressed as a JSON array: `[reaction_id, field, index]` for a reactant, reagent, or solvent (e.g., `["16_1", "reactants", 0]`), and `[reaction_id, "impurities", index, "smiles"]` for an impurity of a preceding step. Following the path by successive indexing of the top-level object (e.g., `data["16_1"]["impurities"][0]["smiles"]`) returns the SMILES string of the parent species. Omitting the final `"smiles"` element instead returns the parent impurity record, which carries its own parent field; repeating this step traces the formation chain of an impurity back to its original source. The field is null for minor_product entries, whose source is the reaction step itself. For five entries of other origins whose source could not be determined from the publication, parent is also null and the note field reads "parent unidentified in source". For model input, the reaction-level reactants_for_impurities field provides the union of all species needed to account for the impurities of a reaction, so that carried-over impurities can be explained from the reactant side alone; 299 additional species are recorded across 128 reactions.

## Dataset Statistics

The dataset contains 229 reactions from 62 publications spanning 2002–2026, with a total of 784 experimentally observed impurity entries. Summary statistics are provided in **Figure 3**.

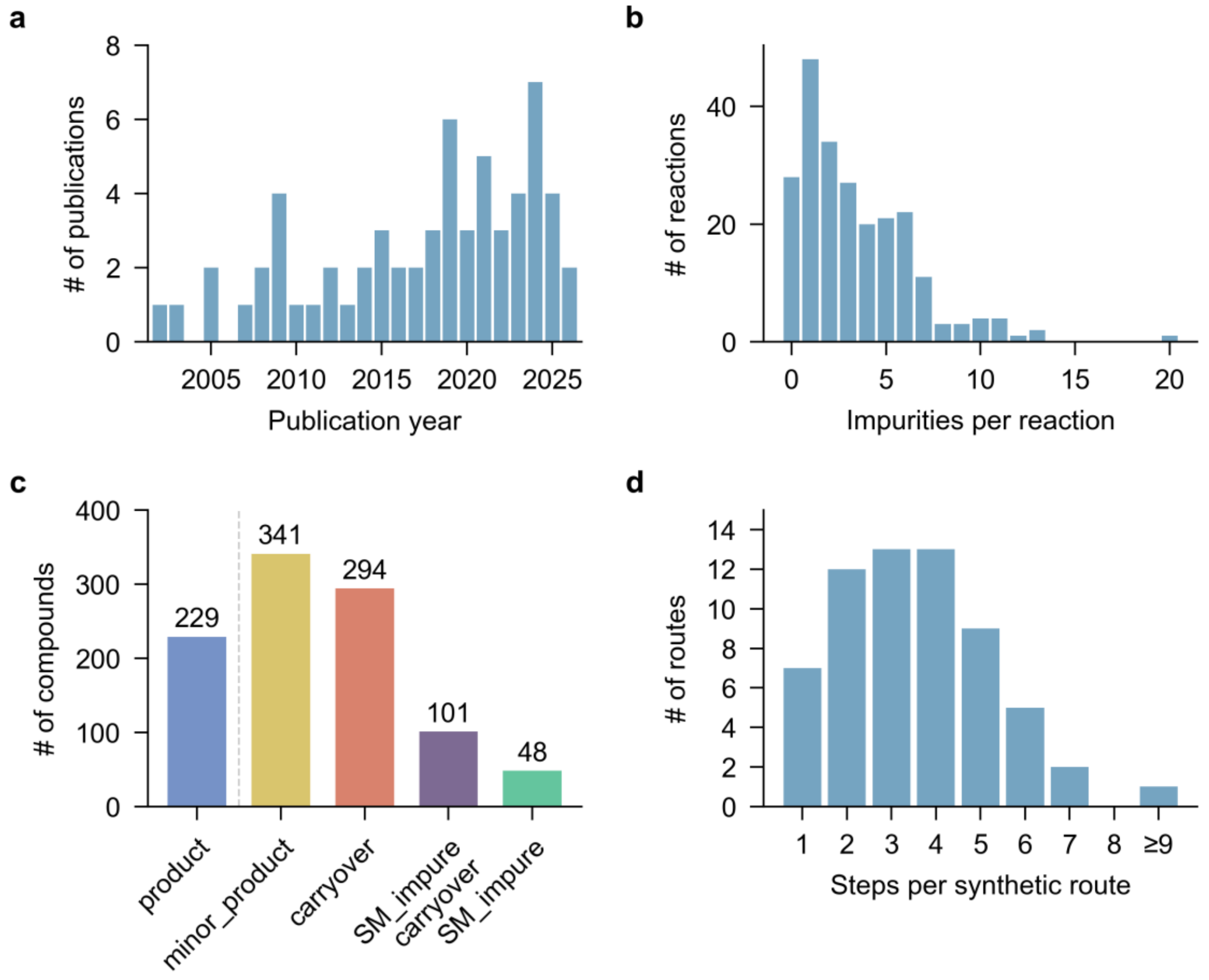

**Figure 3**. Statistical overview of the dataset. (a) Distribution of the 62 source publications by publication year. (b) Distribution of the number of impurity entries per reaction. (c) Number of major products (229) and of impurity entries in each origin category (784 in total). (d) Distribution of synthetic route lengths, i.e., the number of reaction steps recorded per publication.

Reactions with an empty impurity list fall into two groups. Nine reaction steps are intermediate steps of one-pot procedures that were not independently analysed.[27,61,64,65,77,79] The remaining 19 steps (8.3% of 229) are those for which the source publication reports no impurity for that step; they belong to routes whose other steps report impurities and are retained to keep each route complete.[27,28,37,50,57,58,62,64,67] None of them reports complete conversion or a quantitative (100%) yield (isolated yields, where given, range from 44% to 99%), eight of the 19 were telescoped into the following step without isolation or analysis of the intermediate, and only three are reported with 100% HPLC purity of the isolated product, at 77–92% isolated yield. Impurities may therefore have been removed during workup prior to analysis or simply not detected.

A small number of reactions contain two entries with the same structure. These are retained because the source publication reports the compound as arising through two distinct formation pathways, and each entry carries its own origin label and parent assignment.

The dataset spans a broad range of reaction types encountered in pharmaceutical process chemistry, including amide bond formation, reductive amination, N- and O-alkylation, protecting group installation and removal, palladium-catalyzed cross-coupling, and asymmetric transformations. The distribution of impurity counts across reactions is right-skewed, with a median of three impurity entries per reaction (mean 3.4) and a long tail of twelve reactions with ten or more impurities reflecting complex process development scenarios (Figure 3b).

Every reaction was classified with NameRxn[90] (NextMove Software), which assigns a hierarchical reaction name. The recorded major product and each impurity were represented as a reaction whose reactant side comprises the recorded reactants, reagents and solvents together with the starting-material impurities and the parent species of carried-over impurities, and whose product side is the major product or the impurity. Entries that involve no net chemical transformation—unreacted starting material or reagent remaining in the product stream, and impurities carried over unchanged from a preceding step—were nevertheless submitted to NameRxn, and were almost always assigned to the “Resolution” class (typically “Other separation”). In total, 1,013 reactions (229 products and 784 impurities) were classified; the assigned class is stored in the `name_rxn` field of the reaction or impurity, and the exact reaction SMILES submitted to NameRxn in name_rxn_input. We then compared the class distribution of our dataset with those of the pre-processed one-step reaction datasets USPTO (~0.97 M reactions) and Pistachio (~4.3 M reactions).[91]

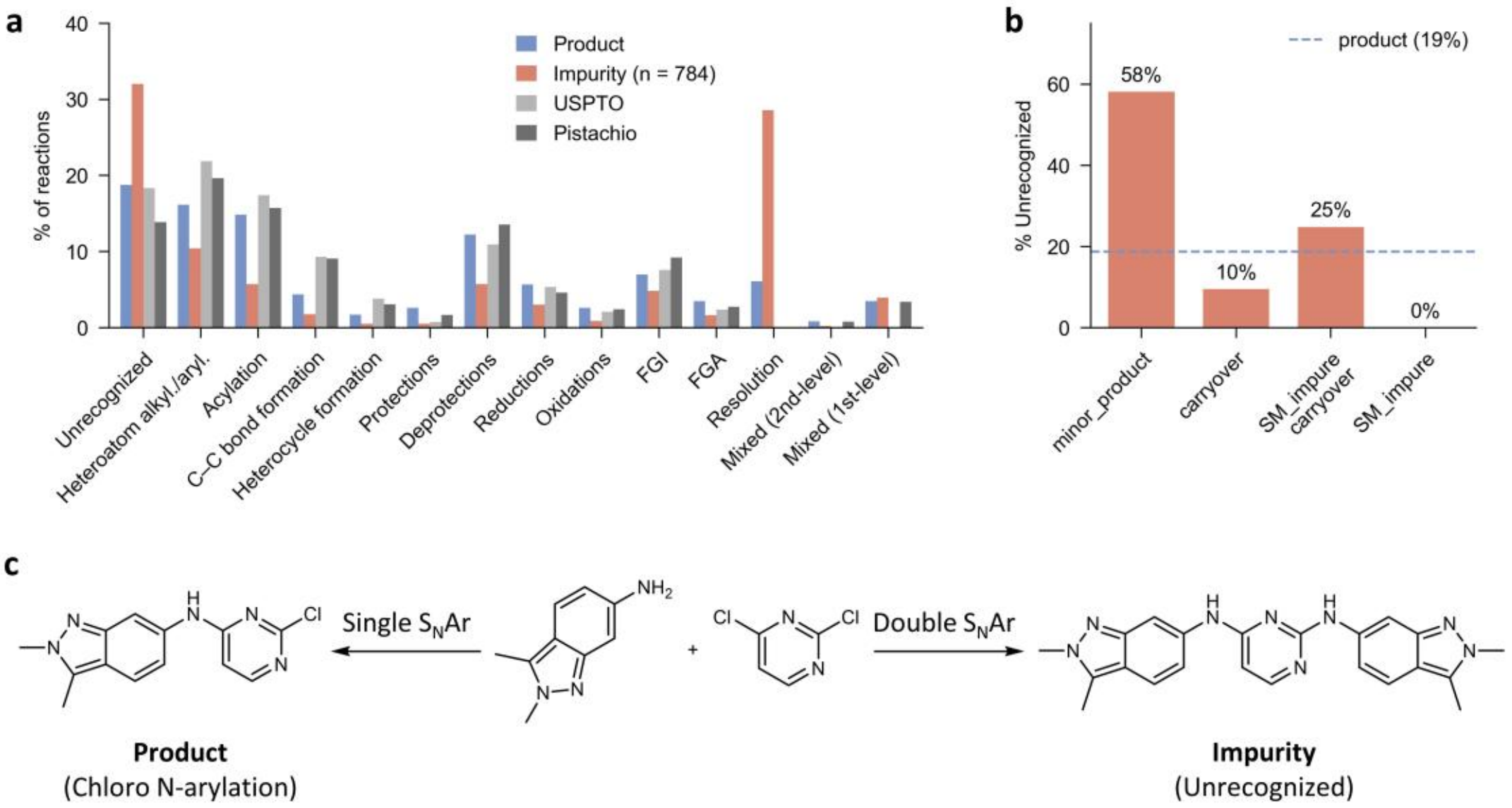

**Figure 4.** Reaction-class analysis of the dataset. (a) Fraction of reactions in each first-level NameRxn class for the major products and the impurities of OPRD-Imp, compared with the USPTO and Pistachio reaction sets. (b) Percentage of “Unrecognized” reactions among impurities, split by annotated origin (n = 341, 294, 101 and 48 for minor_product, carryover, SM_impure_carryover and SM_impure, respectively); the dashed line marks the rate for the major products (19%). (c) A representative unrecognized impurity that is nonetheless a genuine transformation (reaction 4_3): from the same reactants, a single $S_NAr$ gives the intended N-arylation product (left, a named reaction), whereas displacement of the second chloride by a second equivalent of the aminoindazole gives the bis-adduct impurity (right), a net double substitution with no single NameRxn label.

The major products of our impurity dataset populate the classes expected of medicinal- and process-chemistry routes and are almost indistinguishable from USPTO and Pistachio (Figure 4a). The impurities show a different trend: relative to the products, they are enriched mainly in "Resolution" and "Unrecognized". The "Resolution" enrichment, however, is largely explained by unreacted precursors. Of the 224 impurities assigned to this class, 217 (97%) are species already present on the reactant side, i.e., leftover starting material or reagent. The unrecognized fraction is larger for impurities than for the major products (Figure 4a). Splitting the impurities by their annotated origin, we also observed that genuine minor products, the actual products of the step's chemistry, are "Unrecognized" 58% of the time (Figure 4b). The "Unrecognized" label reflects the coverage of the NameRxn ontology as much as the novelty of the chemistry: many impurities are genuine net transformations, including over-reactions, eliminations, and rearrangements, that simply lack a single named-reaction label. Figure 4c shows a representative case. From the same two reactants, a single $S_NAr$ coupling of the aminoindazole with 2,4-dichloropyrimidine gives the intended N-arylation product, while displacement of the second chloride by a second equivalent of the aminoindazole gives the bis-adduct. This is a net double substitution that NameRxn cannot assign to a single class and therefore labels as "Unrecognized".

## Technical Validation

All SMILES strings in the dataset were parsed using RDKit (version 2025.03).[92] Canonical SMILES were generated for every entry, and any entry for which RDKit failed to produce a valid molecular graph was investigated and either corrected or excluded; all SMILES in the released file are canonical. The internal consistency of the released file was then checked programmatically. Every reaction_smiles was verified to reconstruct exactly from the canonical component SMILES of its reactants, reagents, solvents and product, and every name_rxn_input to terminate in the SMILES of the entry it annotates. All 164 next_reaction_id references resolve to keys of the top-level object, and the nine one_pot_group sets are mutually consistent, each member listing the same group. Of the 784 impurity entries, the 438 carrying a parent path all resolve to a SMILES string of a species recorded at the same or an earlier step of the same publication; the remaining entries are the 341 minor_product entries, whose source is the reaction step itself, and the five entries whose source could not be determined from the publication.

In addition to automated SMILES validation, every entry in the dataset was subjected to a multi-round expert review. Each source publication was examined at least three times in total, once during the initial curation and twice more by curators who had not performed the original extraction. Reviewers cross-checked every impurity against the original publication, using the direct evidence quotation recorded during the initial extraction as the primary reference. Entries whose evidence quotation could not be found in the source, or whose structural assignment was chemically implausible, were corrected or removed, yielding the final set of 784 curated impurity entries.

Origin labels were assigned by consensus of the three curators. Where the source publication explicitly stated the origin of an impurity, that assignment was followed. In all other cases, the most plausible origin was determined from the reaction mechanism and recorded only after agreement among all three curators. Stereochemical assignments in the SMILES strings were transcribed as drawn in the source publications and were not independently re-assigned.

## Usage Notes

OPRD-Imp is distributed as a single JSON file (OPRD_imp.json). To reconstruct multi-step synthetic routes, follow the `next_reaction_id` field, which links each reaction to its subsequent step(s) within the same publication. Tracking these links recovers the directed route graph and allows carryover impurities to be tracked across sequential steps. An empty impurities list means that no impurity was reported in the source publication under its described analysis. Because absence of a report is not evidence of absence of impurities, these steps should be treated as unannotated rather than as negative examples (see Dataset

Statistics). Empty lists belonging to intermediate steps of one-pot procedures (identified by the one_pot_group field) are not negative examples, as those steps were not independently analysed. Quantitative information (for example HPLC area percentages) was captured during extraction where reported, but is not included in the released schema because reporting conventions were heterogeneous and frequently incomplete; incorporating normalised quantification is left for future work.

Reactions carried out in a single vessel without isolation of intermediates are identified by the one_pot_group field, an array listing the reaction_id keys of all steps belonging to the same one-pot procedure (e.g., ["16_1", "16_2"]); the field is null for all other reactions, so no string parsing is required. The notes field documents simplified SMILES conventions adopted for species that lack a well-defined molecular structure: palladium on carbon is denoted [C].[Pd], with the carbon support represented by a bare carbon atom, Raney nickel [Al].[Ni], reflecting its Ni–Al alloy composition, activated carbon [C], and palladium(II) chloride [Cl][Pd][Cl]. Molecular hydrogen and nitrogen follow the PubChem notations [HH] and N#N, respectively. Ligands containing ferrocene units, such as 1,1′-bis(diphenylphosphino)ferrocene (dppf), are represented by a disconnected ionic convention, two cyclopentadienylphosphine anions with [Fe+2], as their coordination bonds cannot be expressed in standard SMILES.

To inspect the dataset, we provide an interactive web dashboard, distributed as a single self-contained HTML file (OPRD_imp_dashboard.html) in the same figshare repository[89] that renders each entry as chemical structures, showing its reactants, reagents, solvents, product, and observed impurities with their origin labels, and follows the next_reaction_id links to display multi-step synthetic routes. The dashboard allows users to search and view individual reactions without parsing the raw JSON.

## Data Availability

The OPRD-Imp dataset (OPRD_imp.json) and the interactive dashboard (OPRD_imp_dashboard.html) are available from figshare under a CC BY 4.0 licence at https://doi.org/10.6084/m9.figshare.34075524.[89]

## Code Availability

Reaction-class annotation used NameRxn 3.7.7 (NextMove Software)[90], a commercial tool; the exact reaction SMILES submitted to NameRxn and the returned class labels are stored in the dataset (name_rxn_input and name_rxn fields) so that the annotation can be reproduced by licence holders.

## Funding

This work was supported by the AI Seoul Tech Research Support Program of the Seoul Future Foundation (to H.K.), the National Research Foundation of Korea (NRF) grant funded by the Korean government (MSIT) (RS-2026-25484674, to J.F.J.), and by the Yangyoung Foundation through the Academic Research Support Program (to J.F.J.).

## Author contributions

H.K. conceived and designed the study, designed the data extraction protocol and developed the associated software, and performed the formal analysis and visualization. H.K., S.K., and J.K. carried out the data curation, chemical structure drawing, and validation. T.P. and M.L. reviewed and manually corrected entries in the dataset. S.T. performed the reaction-class analysis. H.K. wrote the original draft, and H.K., S.K., J.K., J.F.J., T.P., M.L., S.T., and C.W.C.

reviewed and edited the manuscript. J.F.J. supervised the project. J.F.J. and J.Y. acquired funding and administered the project. All authors reviewed and approved the final manuscript.

## Competing interests

The authors declare no competing interests.